\documentclass[aps,prb,reprint,superscriptaddress,showpacs,nolongbibliography,balancelastpage]{revtex4-1}

\usepackage{graphicx}   
\usepackage{times}      
\usepackage{amsmath}    
\usepackage{srcltx}     

\newcommand{\HK}{$H_{\mathrm{K}}$}

\newcommand{\CoFe}{Co$_{x}$Fe$_{(1-x)}$}

\newcommand{\tFM}{t$_{\mathrm{M}}$}

\newcommand{\sevenCoFe}{Co$_{70}$Fe$_{30}$}

\newcommand{\Neel}{N\'{e}el}
\newcommand{\cc}{cm$^3$}
\newcommand{\UHA}{[$1\overline{1}0$]}
\newcommand{\UEA}{[$110$]}
\newcommand{\twoCoFeB}{Co$_{20}$Fe$_{60}$B$_{20}$}
\newcommand{\fourCoFeB}{Co$_{40}$Fe$_{40}$B$_{20}$}
\newcommand{\sixCoFeB}{Co$_{60}$Fe$_{20}$B$_{20}$}
\newcommand{\sevenCoFeB}{Co$_{68}$Fe$_{22}$B$_{10}$}

\newcommand{\degreesC}{$^{\circ}$C}
\newcommand{\degrees}{$^{\circ}$}
\newcommand{\CoFeBtwo}{(CoFe)$_{80}$B$_{20}$}

\newcommand{\manuscript}{Brief Report}

\begin{document}

\title{Origin of in-plane uniaxial magnetic anisotropy in CoFeB amorphous ferromagnetic thin-films}

\author{A.T.~Hindmarch}
\email[email:~]{aidan.hindmarch@nottingham.ac.uk}
\affiliation{School of Physics \&\ Astronomy, University of
Nottingham, Nottingham, NG7 2RD, United Kingdom}
\affiliation{School of Physics \&\ Astronomy, University of Leeds, Leeds, LS2 9JT, United Kingdom}

\author{A.W.~Rushforth}
\author{R.P.~Campion}
\affiliation{School of Physics \&\ Astronomy, University of
Nottingham, Nottingham, NG7 2RD, United Kingdom}

\author{C.H.~Marrows}
\affiliation{School of Physics \&\ Astronomy, University of Leeds, Leeds, LS2 9JT, United Kingdom}

\author{B.L.~Gallagher}
\affiliation{School of Physics \&\ Astronomy, University of
Nottingham, Nottingham, NG7 2RD, United Kingdom}

\begin{abstract}
Describing the origin of uniaxial magnetic anisotropy (UMA) is generally problematic in systems other than single crystals. We demonstrate an in-plane UMA in amorphous CoFeB films on GaAs(001) which has the expected symmetry of the interface anisotropy in ferromagnetic films on GaAs(001), but strength which is independent of, rather than in inverse proportion to, the film thickness. We show that this volume UMA is consistent with a bond-orientational anisotropy, which propagates the interface-induced UMA through the thickness of the amorphous film. It is explained how, in general, this mechanism may describe the origin of in-plane UMAs in amorphous ferromagnetic films.
\end{abstract}


\date{\today}
\maketitle


Magnetic materials possessing a uniaxial magnetic anisotropy (UMA) find many important applications in fields such as information storage and magnetic field sensors. For example, materials possessing a uniaxial perpendicular to the plane magnetic anisotropy (PMA) have been used in magnetic recording media such as hard disk drives (HDD), while ferromagnetic (FM) thin-films with an in-plane UMA component, particularly CoFe-based alloys, are becoming increasingly important for applications in the rapidly developing field of spintronics. In particular, CoFeB thin-films are routinely used in a range of studies including tunneling magnetoresistance~\cite{Djayaprawira2005}\ and current induced magnetization switching~\cite{Kubota2008}, and are utilized in commercial applications such as HDD read-heads and magnetic random access memories. Therefore, proper understanding of the microstructural origins of the anisotropy in this system is of great technological, in additional to fundamental, importance.

In magnetic thin-films, the `effective' magnetic anisotropy constants ($K^\mathrm{eff}$) are generally described in terms of volume ($K^{\mathrm{vol}}$) and interface ($K^{\mathrm{int}}$) contributions as
\begin{equation*}\label{eqK}
K_{\mathrm{a}}^\mathrm{eff}=K_\mathrm{a}^{\mathrm{vol}}+K_\mathrm{a}^{\mathrm{int}}/t_\mathrm{M},
\end{equation*}
where $\mathrm{a}=\mathrm{U}(\bot)$, $1$, $2$, etc.\ describe uniaxial (perpendicular), first and second order cubic anisotropies, and so forth, and $t_\mathrm{M}$ is the magnetic film thickness. 
Magnetocrystalline anisotropy, having its origin in spin-orbit coupling and reflecting the crystal symmetry, often accounts for the volume contribution in crystalline materials, whereas strain or spin-orbit interactions at the interface can account for the interface contribution.


It is also possible for an amorphous material to possess a volume UMA, although it can be unclear exactly what the microstructural origin of such a contribution can be. For example, certain rare earth - transition metal intermetallic (RE-TM) compounds possess a PMA which is a volume contribution. The mechanism for the volume PMA in amorphous RE-TMs has been extensively debated, with `bond-orientational' anisotropy (BOA) emerging as the most commonly suggested mechanism~\cite{Robinson1989,Yan1991,Harris1992}. BOA refers to a medium-to-long-range \textit{microstructural}\ anisotropy corresponding to orientational correlation of anisotropic local coordination polyhedra~\cite{Suzuki1987,Yan1991}. The \Neel --Taniguchi (N--T)~\cite{Neel1954,*Taniguchi1955}\ directional pair-ordering model is also frequently suggested; within this model PMA is introduced via anisotropic dipole-like coupling between individual atom-pairs --- anisotropic \textit{chemical}\ ordering of near-neighbor atoms in randomly oriented coordination polyhedra results.


In this \manuscript\ we show that the in-plane UMA observed in prototypical amorphous CoFeB films on GaAs, previously assumed to be purely an interface contribution, is, in fact a volume anisotropy that is seeded by an interface interaction during the growth. By studying the effects of varying the film thickness and composition, and by applying elastic strain, we show that the microstructural origin of the volume UMA in this system is consistent only with the BOA model. This should also be the origin of the UMA in such films, what-ever is the azimuthal symmetry breaking mechanism during the film deposition - be it an applied magnetic field, interface interaction, or oblique deposition geometry.

A GaAs(001) epilayer was deposited onto a 2$^{\prime\prime}$ diameter GaAs(001) wafer by III-V molecular beam epitaxy, and capped with arsenic to prevent oxidation. The wafer was diced into $8\times 8$~mm$^2$ pieces and mounted into an ultra-high vacuum (UHV) deposition system. After thermally desorbing the As-passivation and allowing the epilayers to cool to 45~\degreesC\ under UHV, CoFeB[\tFM]/Ta[2~nm] and CoFe[\tFM]/Ta[2~nm] films were deposited by dc magnetron sputtering at \textit{normal incidence}: no magnetic field was applied during deposition, and the residual (toroidal) field from the sources was less that 0.1~Oe at the sample position. The \tFM\ were in the range 3.5--20.0~nm, and alloy sputter targets of the stated compositions were used. \CoFeBtwo\ (at.~\%) films are amorphous~\cite{Hindmarch2008,Platt2001} whilst \sevenCoFe\ films grow epitaxially with bcc structure and (001) orientation~\cite{Hindmarch2010}. Magnetic characterization of the samples was performed at low temperature using vibrating sample (VSM) (10~K) and SQUID (25 -- 250~K) magnetometries, following initial characterization at room temperature using magneto-optical Kerr-effect magnetometry. Piezo-actuated devices were fabricated in order to apply elastic strain: the GaAs substrate was partially etched away from $\sim 3\times2$~mm$^2$ pieces of the \fourCoFeB [20~nm]/GaAs(001) sample to leave $\sim 150 \mu$m GaAs, and were bonded onto commercial piezo-transducers. Details of the fabrication of such devices using (Ga,Mn)As epilayers may be found in reference~\onlinecite{Rushforth2008}. The transducer is aligned to $\sim 5$~\degrees\ of the GaAs\UHA\ direction, and applying a voltage of $+150$~V ($-50$~V) results in uniaxial tensile (compressive) strain nominally along the GaAs\UHA\ direction, with concomitant compressive (tensile) strain along \UEA. Between $V=-50$~V and $+150$~V, strain is $\sim 4\times 10^{-4}$ at 150~K from strain-gauge measurements.


First we present the experimental evidence to establish that the UMA is a volume contribution and that it is seeded by the interface. Inducing in-plane \textit{volume}\ UMA during film deposition requires azimuthal symmetry to be broken: typically achieved by applying an in-plane magnetic field~\cite{Raanaei2009}, or using a deposition geometry whereby the atomic flux impinges at an angle to the substrate normal~\cite{Gonzalez-Guerrero2007}. In the absence of such symmetry breaking, random local magnetic anisotropy results, due to short-range ordering in the amorphous FM. Alternatively, and of particular relevance in spintronics, in-plane UMA is found in \textit{epitaxial}\ FM metal films deposited onto the (001) surface of III-V semiconductors~\cite{Krebs1987,Thomas2003,Hindmarch2010}. In these systems an \textit{interface}\ UMA arises, with uniaxial easy axis (UEA) along, e.g., the \UEA\ direction of GaAs(001). Amorphous CoFeB films on GaAs(001) also exhibit interface-induced UMA~\cite{Hindmarch2008}: this interface interaction results in azimuthal symmetry breaking during growth. 

\begin{figure}
\includegraphics[width=0.8\linewidth]{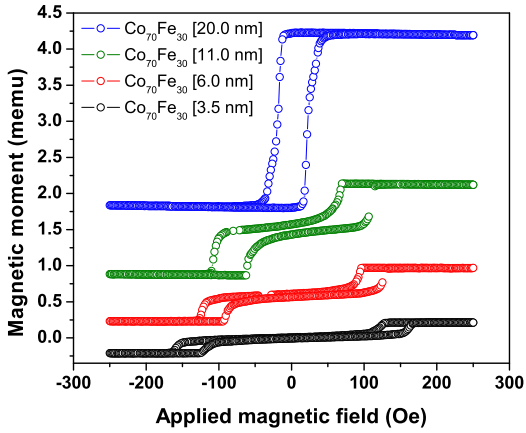}\\
\caption{(Color online) VSM hysteresis loops for prototype sputtered epitaxial \sevenCoFe\ films on GaAs(001), measured along GaAs\UHA\ at 10K. The effective uniaxial anisotropy dies off with increasing film thickness, as anticipated. Data are offset for clarity.}\label{figCoFe}
\end{figure}

Figure~\ref{figCoFe}\ shows VSM hysteresis loops along the uniaxial hard-axis (UHA) for prototypical epitaxial \sevenCoFe\ films on GaAs(001). The UEA in \sevenCoFe\ films are along GaAs\UEA\ and UHA along \UHA: as expected for UMA arising due to the interface interaction with GaAs(001). For thinner films, characteristic two-stage magnetization reversal is observed due to competition between cubic (volume) and uniaxial (interface) anisotropy contributions of similar strength~\cite{Krebs1987,Bianco2008,Hindmarch2010}: in the thickest film the interfacial UMA is significantly weaker than the volume cubic anisotropy. Conventional behavior in these epitaxial films confirms that the UMA is attributable to the interface interaction with GaAs(001).


\begin{figure}
\includegraphics[width=\linewidth]{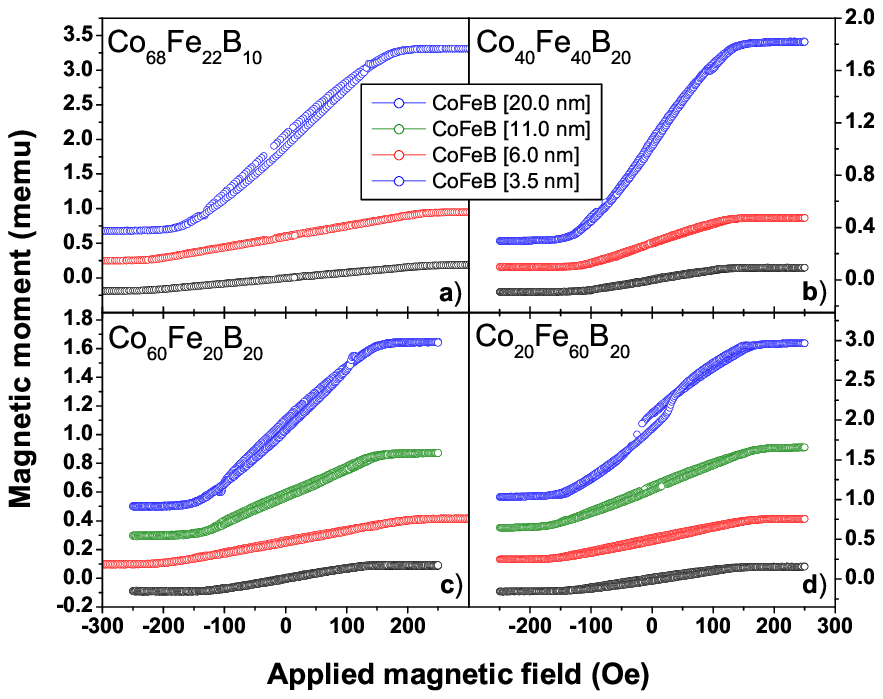}\\
\caption{(Color online) VSM hysteresis loops for sputtered CoFeB films on GaAs(001), measured along GaAs\UHA\ at 10~K. CoFeB alloy compositions are a) \sevenCoFeB, b) \fourCoFeB, c) \sixCoFeB, and d) \twoCoFeB. Data are offset for clarity.}\label{figCoFeB}
\end{figure}

VSM hysteresis loops for CoFeB films, of various thickness and composition, on GaAs(001), are shown in figure~\ref{figCoFeB}. In all cases, strong UMA is observed with UEA along the GaAs\UEA\ and UHA along \UHA; consistent with the direction of interfacial UMA in FM/GaAs(001), e.g., refs~\onlinecite{Krebs1987,Thomas2003,Hindmarch2008}, and figure~\ref{figCoFe}. All samples exhibit anhysteretic hard-axis reversal, with small departures due to slight misalignment ($\sim 2$~\degrees) between the film-plane and applied measurement field. The anisotropy field $H_\mathrm{K}=2K_{\mathrm{U}}^{\mathrm{eff}}/M_\mathrm{S}$ provides a direct measure of the effective UMA constant, $K_{\mathrm{U}}^{\mathrm{eff}}$, for a FM with saturation magnetization $M_\mathrm{S}$: amorphous CoFeB films deposited in an applied magnetic field typically have \HK~$\sim 40$~Oe (corresponding to $K_{\mathrm{U}}^{\mathrm{eff}}\sim 2\times 10^4$~erg/\cc), e.g., ref.~\onlinecite{Kirk2009},  and \HK~$\sim 70$~Oe has been found in Co$_{68}$Fe$_{24}$Zr$_{8}$~\cite{Raanaei2009}. Thus, despite the absence of an applied magnetic field during film deposition, we find a significantly \textit{enhanced}\ UMA, with \HK$\sim 150$~Oe ($K_{\mathrm{U}}^{\mathrm{eff}}\sim 8\times 10^4$~erg/\cc) in amorphous CoFeB 
on GaAs(001). UMA of such strength has previously been observed in amorphous CoFeB films deposited at oblique-incidence~\cite{Gonzalez-Guerrero2007}, confirming that the UMA should be related in some way to anisotropy in the film microstructure~\cite{Park1995}. However, our films were deposited at \textit{normal incidence}.

\textit{Most importantly}, we observe a surprising thickness dependence of UMA for \CoFeBtwo\ films on GaAs(001). The UMA is \textit{independent of, rather than inversely proportional to}\, the thickness of the amorphous FM film.

\begin{figure}
\includegraphics[width=0.9\linewidth]{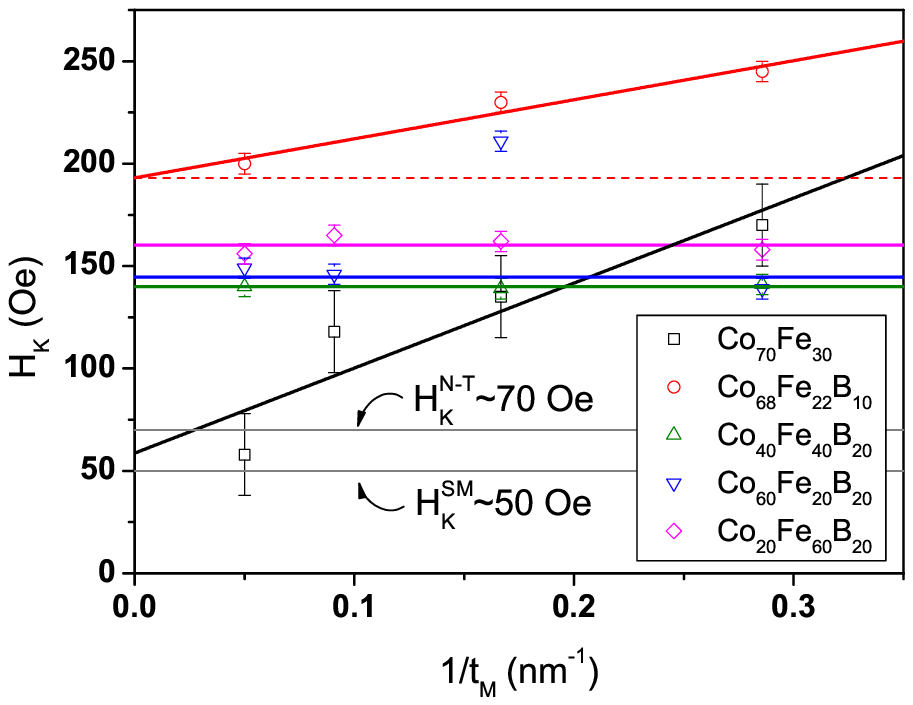}\\
\caption{(Color online) Inverse thickness dependence of the uniaxial anisotropy field, \HK, in epitaxial \sevenCoFe, and in amorphous \sevenCoFeB, \fourCoFeB, \twoCoFeB, and \sixCoFeB. Thick lines are guides to the eye and the dashed line represents the volume contribution to the anisotropy in \sevenCoFeB.}\label{figKU}
\end{figure}

Figure~\ref{figKU}\ summarizes the thickness dependence of \HK\ from \sevenCoFe\ and CoFeB films, from figures~\ref{figCoFe}\ and \ref{figCoFeB}. For epitaxial \sevenCoFe, \HK\ is roughly in inverse proportion to thickness, characteristic of an interface anisotropy~\cite{Note1}%
. However, for \CoFeBtwo\ films, \HK\ $\sim 150$~Oe, and shows no significant thickness (or composition) dependence; demonstrating that the UMA {is}, in-fact, a \textit{volume}\ anisotropy rather than the \textit{interface}\ anisotropy which one may na\"{i}vely predict. 

We now consider the mechanisms by which we may anticipate a volume UMA to arise in an amorphous film as the result of an interfacial interaction. Theoretical work by Fu and Mansuripur~\cite{Fu1992}, considering the orientation of pair-bonds in an amorphous film (thus encapsulating N--T and/or BOA models), demonstrated that a volume UMA results due to the Boltzmann distribution of bond-orientations: this anisotropic distribution being induced by the initial magnetization of the growing film. Whilst ref.~\onlinecite{Fu1992}\ specifically considered volume PMA in amorphous ferrimagnetic RE-TM films, the same principle (with the exception that the sign of the inter-sublattice coupling makes PMA unfavorable due to shape anisotropy) holds for in-plane UMA in any amorphous 
FM film. Thus, whichever microstructural mechanism generally provides volume UMA in CoFeB, the \textit{interface}\ anisotropy due to the GaAs(001) substrate \textit{should}\ seed a \textit{volume}\ UMA in the amorphous CoFeB film with UEA along the GaAs\UEA\ direction; as demonstrated in figures~\ref{figCoFeB}\ and \ref{figKU}.

Given that either of the proposed microstructural mechanisms for UMA in amorphous FM films may be expected to produce a volume UMA in CoFeB on GaAs(001), we look now to distinguish the N--T and BOA mechanisms by other means. We note that most techniques capable of providing `direct' microstructural information on amorphous ultrathin-films, e.g., diffraction or spectroscopy methods, yield \textit{ensemble averaged, one-dimensional}\ information (radial pair-distribution functions, etc.)~\cite{Robinson1989,Yan1991,Harris1992,Sheng2006,Kirk2009,Hirata2011}: these techniques are typically unable to clearly distinguish the microstructural origin of UMA, particularly in CoFe-based amorphous alloys.


In many RE-TM alloys the PMA is found to be strongly composition dependent: however, this is dominated by the large orbital magnetic moment of the 4$f$ shell in the REs (with the exception of Gd). As Fe and Co are similar in terms of atomic volume, bonding coordination, and (partially quenched) orbital magnetic moment --- certainly not the case for TMs and REs --- one may expect that the composition dependence of the UMA in CoFe-based amorphous alloys may allow us to differentiate N--T and BOA mechanisms.

We may estimate the UMA which may be anticipated in the case of the N--T mechanism by considering, as an analog to CoFeB, random polycrystalline \CoFe\ alloys: the volume UMA then is $K_\mathrm{U}^\mathrm{vol, N-T}=a x^2(1-x)^2(T_\mathrm{C}-T_\mathrm{d})$~\cite{Srivastava1977}, where $T_\mathrm{C}\approx 1300$~K is the Curie temperature and $T_\mathrm{d}\approx 320$~K is the deposition temperature. Taking the weakly composition dependent empirical prefactor $a \sim 600$~erg/\cc$\cdot$K~(ref.~\onlinecite{Srivastava1977}), and the measured $M_\mathrm{S}\approx1100$~emu/\cc\ for \fourCoFeB, results in UMA with $K_\mathrm{U}^\mathrm{vol, N-T}\sim 3.5\times10^4$~erg/\cc\ (giving $H_\mathrm{K}^{\mathrm{N-T}}\sim 70$~Oe) for equal Co:Fe ratio. Given the approximations employed, this value is reasonably close to our experimental result. However, strong compositional dependence of the UMA is expected within the N--T model: which we do not observe in our films. Maximum $K_\mathrm{U}^\mathrm{vol, N-T}$ should be observed for equal Co:Fe composition, with the UMA diminishing rapidly as composition, $x$, varies. However, we find very weak composition dependence in \CoFeBtwo\ alloys (figures~\ref{figCoFeB}\ and \ref{figKU}), incompatible with the strong composition-dependence expected for N--T pair-anisotropy.


Within the BOA model bonding is not specifically chemically dependent; one may expect negligible microstructural variation upon substituting similar atomic species such as Co and Fe. The UMA is the result of a combination of medium-range `anelastic' microscopic and long-range elastic macroscopic strains, produced due to the anisotropic orientation of local coordination polyhedra: anelastic strain is temporarily non-recoverable, e.g., after removing stress, but, in many cases, may be recovered by annealing. Anelastic microscopic strain $\epsilon_\mathrm{a}$ may be up to $\sim5$~times greater than the resulting elastic macroscopic strain $\epsilon$ in both FM TM-metalloid alloys ~\cite{Suzuki1987}\ and amorphous RE-TM ferrimagnets~\cite{Yan1991}. A weak change in UMA may be expected with varying Co:Fe composition within the BOA picture, due to the composition-dependent magnetostriction in amorphous \CoFeBtwo\ alloys.

\begin{figure}
\includegraphics[width=0.63\linewidth]{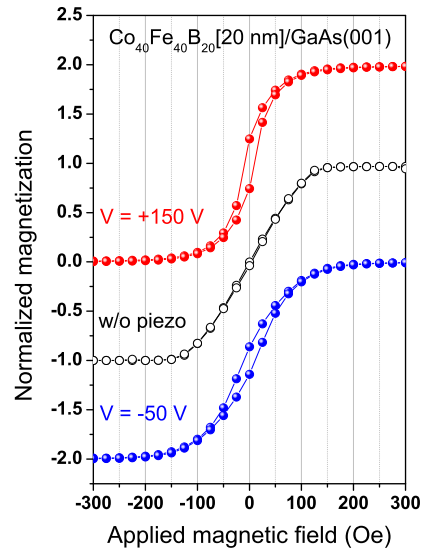}\\
\caption{(Color online) SQUID hysteresis loops for amorphous \fourCoFeB[20~nm] on GaAs(001), measured along GaAs\UHA\ at 150K.  Loops for piezo-actuated devices with voltages +150~V and -50~V, and without piezo-transducer, are shown, offset for clarity.}\label{figPiezo}
\end{figure}

In \textit{polycrystalline}\ FMs deposited in an applied magnetic field, \textit{elastic}\ strain-magnetostriction (SM) anisotropy, caused by the inability of the FM film to deform under magnetization rotation due to the constraint at the substrate, may result in a volume UMA. The anisotropy constant in this case is commonly approximated as $K_\mathrm{U}^{\mathrm{vol, SM}} \sim \frac{3}{2} Y \lambda^2$ [refs.~\onlinecite{Robinson1962,West1964}], where $\lambda$ is the (polycrystal averaged) Joule magnetostriction and $Y$ the Young's modulus. Whilst the \textit{elastic}\ SM mechanism \textit{alone}\ should not arise in an amorphous FM due to the lack of long-range structural coordination, it is instructive to determine the macroscopic elastic strain-response of amorphous \fourCoFeB\ on GaAs(001).

Hysteresis loops for a piezo-actuated device are shown in figure~\ref{figPiezo}: the magnetic field is applied along the nominal UHA, and applied \textit{macroscopic}\ strain causes UHA to rotate slightly from the field direction~\cite{Rushforth2008}. There is a clear change in UMA due to the applied strain, $\Delta K_{\mathrm{U}}^{\mathrm{vol}}$, which may be determined from the difference in the integrals $\int_{0}^{M_{\mathrm{S}}} H(M) dM$ for the hysteresis loops in figure~\ref{figPiezo}: from which $\Delta K_{\mathrm{U}}^{\mathrm{vol}}\approx 2.2\times10^4$~erg/\cc. The change in UMA under macroscopic stress, $\sigma$, is $\Delta K_{\mathrm{U}}^\mathrm{vol}(\sigma)=\lambda \sigma$: thus
under applied {macroscopic}\ strain $\lambda=\Delta K_{\mathrm{U}}^\mathrm{vol}(\epsilon)/Y\epsilon$. Taking $Y=162$~GPa~\cite{Barandiaran2003}, we find $\lambda \approx 3.5\times 10^{-5}$: consistent with what may be anticipated for \CoFeBtwo\ alloy films~\cite{Platt2001,Gonzalez-Guerrero2007}. Thus, as $K_\mathrm{U}^{\mathrm{vol, SM}}\sim 3\Delta K_{\mathrm{U}}^{\mathrm{vol}}(\epsilon)^2/2Y\epsilon^2  \sim 3\times 10^4$~erg/\cc,
one would expect solely \textit{elastic}\ SM in \CoFeBtwo\ may produce UMA with $H_{\mathrm{K}}\sim 50$~Oe. Thus, this is consistent with the UMA with \HK$\sim150$~Oe found in \CoFeBtwo\ on GaAs(001) being due to the combination of \textit{elastic}\ long-, and stronger \textit{anelastic}, medium-range strains which arise via the BOA mechanism.


In films with lower B concentration ($\sim 4$~\%) the bcc-like CoFe structure has been shown to be distorted due to the glass-forming metalloid additive in combination with oblique deposition, resulting in large \HK$\sim 500$~Oe~\cite{Hashimoto2008}. As the local atomic structure in \sevenCoFeB\ films should also be more closely related to a (distorted) bcc crystal than that for \CoFeBtwo~\cite{Platt2001,Sheng2006}, the locally bcc-like structure may result in the observed weak residual \textit{interfacial}\ UMA. Anisotropic bonding is intrinsic to such a distorted bcc-like local structure: the strong \textit{volume}\ UMA may also be anticipated within the BOA mechanism due to the larger $Y$ and $\lambda$ in \sevenCoFeB\ [and (CoFe)$_{96}$B$_4$] over \CoFeBtwo~\cite{Platt2001,Barandiaran2003}. Thus, as in figure~\ref{figKU}, the UMA in disordered \sevenCoFeB\ comprises an admixture of thickness-dependent and -independent UMA terms found, respectively, in crystalline and amorphous FM films.


Finally, we comment on why the \textit{volume}\ UMA due to oblique deposition \textit{or}\ interface interaction dominates over that due to applied magnetic field during deposition~\cite{Gonzalez-Guerrero2007,Hindmarch2008,Kirk2009}. During the initial stages of growth, a magnetic field alone may induce BOA by aligning the magnetic moments $\mu$, and hence microstructure, of isolated superparamagnetic coordination clusters, typically with $\sim 10$ atoms~\cite{Sheng2006}\ and hence $\mu \sim 20$~$\mu_\mathrm{B}$. The degree of orientational alignment along the applied field direction follows the Langevin function and, for such clusters, should be relatively small due to the $\sim 100$s~Oe typical applied field~\cite{Kirk2009,Raanaei2009,Hindmarch2008}; resulting in a weak initial BOA due to the applied magnetic field, which is then propagated through the growing amorphous FM film.

%




In conclusion, we have demonstrated that an in-plane \textit{volume}\ uniaxial magnetic anisotropy may be produced in prototypical amorphous CoFeB thin-films on GaAs(001) as a consequence of the \textit{interface}\ interaction. The uniaxial magnetic anisotropy has characteristics only consistent with the bond-orientational anisotropy: we suggest that this microstructural mechanism may also be the means by which in-plane uniaxial magnetic anisotropies arise in ferromagnetic amorphous alloy films deposited in a 
magnetic field or at oblique incidence.

The authors acknowledge support from EPSRC Grant Refs.\ EP/H003487/1 and EP/E016413/1, and EU Grant No.\ 214499 NAMASTE. We are grateful to C.T.~Foxon, G.~Burnell, D.~Ciudad, J.S.~Claydon, and M.~Ali for useful discussions.



\begin{thebibliography}{28}%
\makeatletter
\providecommand \@ifxundefined [1]{%
 \@ifx{#1\undefined}
}%
\providecommand \@ifnum [1]{%
 \ifnum #1\expandafter \@firstoftwo
 \else \expandafter \@secondoftwo
 \fi
}%
\providecommand \@ifx [1]{%
 \ifx #1\expandafter \@firstoftwo
 \else \expandafter \@secondoftwo
 \fi
}%
\providecommand \natexlab [1]{#1}%
\providecommand \enquote  [1]{``#1''}%
\providecommand \bibnamefont  [1]{#1}%
\providecommand \bibfnamefont [1]{#1}%
\providecommand \citenamefont [1]{#1}%
\providecommand \href@noop [0]{\@secondoftwo}%
\providecommand \href [0]{\begingroup \@sanitize@url \@href}%
\providecommand \@href[1]{\@@startlink{#1}\@@href}%
\providecommand \@@href[1]{\endgroup#1\@@endlink}%
\providecommand \@sanitize@url [0]{\catcode `\\12\catcode `\$12\catcode
  `\&12\catcode `\#12\catcode `\^12\catcode `\_12\catcode `\%12\relax}%
\providecommand \@@startlink[1]{}%
\providecommand \@@endlink[0]{}%
\providecommand \url  [0]{\begingroup\@sanitize@url \@url }%
\providecommand \@url [1]{\endgroup\@href {#1}{\urlprefix }}%
\providecommand \urlprefix  [0]{URL }%
\providecommand \Eprint [0]{\href }%
\@ifxundefined \urlstyle {%
  \providecommand \doi  [0]{\begingroup \@sanitize@url \@doi}%
  \providecommand \@doi [1]{\endgroup \@@startlink {\doibase
  #1}doi:\discretionary {}{}{}#1\@@endlink }%
}{%
  \providecommand \doi  [0]{doi:\discretionary{}{}{}\begingroup
  \urlstyle{rm}\Url }%
}%
\providecommand \doibase [0]{http://dx.doi.org/}%
\providecommand \Doi [0]{\begingroup \@sanitize@url \@Doi }%
\providecommand \@Doi  [1]{\endgroup\@@startlink{\doibase#1}\@@Doi}%
\providecommand \@@Doi [1]{#1\@@endlink}%
\providecommand \selectlanguage [0]{\@gobble}%
\providecommand \bibinfo  [0]{\@secondoftwo}%
\providecommand \bibfield  [0]{\@secondoftwo}%
\providecommand \translation [1]{[#1]}%
\providecommand \BibitemOpen [0]{}%
\providecommand \bibitemStop [0]{}%
\providecommand \bibitemNoStop [0]{.\EOS\space}%
\providecommand \EOS [0]{\spacefactor3000\relax}%
\providecommand \BibitemShut  [1]{\csname bibitem#1\endcsname}%
\bibitem [{\citenamefont {Djayaprawira}\ \emph {et~al.}(2005)\citenamefont
  {Djayaprawira} \emph {et~al.}}]{Djayaprawira2005}%
  \BibitemOpen
  \bibfield  {author} {\bibinfo {author} {\bibfnamefont {D.~D.}\ \bibnamefont
  {Djayaprawira}} \emph {et~al.},\ }\href@noop {} {\bibfield  {journal}
  {\bibinfo  {journal} {Appl. Phys. Lett.},\ }\textbf {\bibinfo {volume}
  {86}},\ \bibinfo {pages} {092502} (\bibinfo {year} {2005})}\BibitemShut
  {NoStop}%
\bibitem [{\citenamefont {Kubota}\ \emph {et~al.}(2008)\citenamefont {Kubota}
  \emph {et~al.}}]{Kubota2008}%
  \BibitemOpen
  \bibfield  {author} {\bibinfo {author} {\bibnamefont {H.}}\ {\bibnamefont {Kubota}} \emph
  {et~al.},\ }\href@noop {} {\bibfield  {journal} {\bibinfo  {journal} {Nat.
  Phys.},\ }\textbf {\bibinfo {volume} {4}},\ \bibinfo {pages} {37} (\bibinfo
  {year} {2008})}\BibitemShut {NoStop}%
\bibitem [{\citenamefont {Robinson}\ \emph {et~al.}(1989)\citenamefont
  {Robinson}, \citenamefont {Samant},\ and\ \citenamefont
  {Marinero}}]{Robinson1989}%
  \BibitemOpen
  \bibfield  {author} {\bibinfo {author} {\bibfnamefont {C.~J.}\ \bibnamefont
  {Robinson}}, \bibinfo {author} {\bibfnamefont {M.~G.}\ \bibnamefont
  {Samant}}, \ and\ \bibinfo {author} {\bibfnamefont {E.~E.}\ \bibnamefont
  {Marinero}},\ }\href@noop {} {\bibfield  {journal} {\bibinfo  {journal} {Appl
  Phys A},\ }\textbf {\bibinfo {volume} {49}},\ \bibinfo {pages} {619}
  (\bibinfo {year} {1989})}\BibitemShut {NoStop}%
\bibitem [{\citenamefont {Yan}\ \emph {et~al.}(1991)\citenamefont {Yan} \emph
  {et~al.}}]{Yan1991}%
  \BibitemOpen
  \bibfield  {author} {\bibinfo {author} {\bibfnamefont {X.}~\bibnamefont
  {Yan}} \emph {et~al.},\ }\href@noop {} {\bibfield  {journal} {\bibinfo
  {journal} {Phys. Rev. B},\ }\textbf {\bibinfo {volume} {43}},\ \bibinfo
  {pages} {9300} (\bibinfo {year} {1991})}\BibitemShut {NoStop}%
\bibitem [{\citenamefont {Harris}\ \emph {et~al.}(1992)\citenamefont {Harris}
  \emph {et~al.}}]{Harris1992}%
  \BibitemOpen
  \bibfield  {author} {\bibinfo {author} {\bibfnamefont {V.~G.}\ \bibnamefont
  {Harris}} \emph {et~al.},\ }\href@noop {} {\bibfield  {journal} {\bibinfo
  {journal} {Phys. Rev. Lett.},\ }\textbf {\bibinfo {volume} {69}},\ \bibinfo
  {pages} {1939} (\bibinfo {year} {1992})}\BibitemShut {NoStop}%
\bibitem [{\citenamefont {Suzuki}\ \emph {et~al.}(1987)\citenamefont {Suzuki},
  \citenamefont {Haimovich},\ and\ \citenamefont {Egami}}]{Suzuki1987}%
  \BibitemOpen
  \bibfield  {author} {\bibinfo {author} {\bibfnamefont {Y.}~\bibnamefont
  {Suzuki}}, \bibinfo {author} {\bibfnamefont {J.}~\bibnamefont {Haimovich}}, \
  and\ \bibinfo {author} {\bibfnamefont {T.}~\bibnamefont {Egami}},\
  }\href@noop {} {\bibfield  {journal} {\bibinfo  {journal} {Phys. Rev. B},\
  }\textbf {\bibinfo {volume} {35}},\ \bibinfo {pages} {2162} (\bibinfo {year}
  {1987})}\BibitemShut {NoStop}%
\bibitem [{\citenamefont {N{\'{e}}el}(1954)}]{Neel1954}%
  \BibitemOpen
  \bibfield  {author} {\bibinfo {author} {\bibfnamefont {L.}~\bibnamefont
  {N{\'{e}}el}},\ }\Doi {10.1051/jphysrad:01954001504022500} {\bibfield
  {journal} {\bibinfo  {journal} {J. Phys. Radium},\ }\textbf {\bibinfo
  {volume} {15}},\ \bibinfo {pages} {225} (\bibinfo {year} {1954})}\BibitemShut
  {NoStop}%
\bibitem [{\citenamefont {Taniguchi}(1955)}]{Taniguchi1955}%
  \BibitemOpen
  \bibfield  {author} {\bibinfo {author} {\bibfnamefont {S.}~\bibnamefont
  {Taniguchi}},\ }\href {http://hdl.handle.net/10097/26705} {\bibfield
  {journal} {\bibinfo  {journal} {Sci. Rep. Res. Inst. Tohoku Univ. A},\
  }\textbf {\bibinfo {volume} {7}},\ \bibinfo {pages} {269} (\bibinfo {year}
  {1955})}\BibitemShut {NoStop}%
\bibitem [{\citenamefont {Hindmarch}\ \emph {et~al.}(2008)\citenamefont
  {Hindmarch} \emph {et~al.}}]{Hindmarch2008}%
  \BibitemOpen
  \bibfield  {author} {\bibinfo {author} {\bibfnamefont {A.~T.}\ \bibnamefont
  {Hindmarch}} \emph {et~al.},\ }\href@noop {} {\bibfield  {journal} {\bibinfo
  {journal} {Phys. Rev. Lett.},\ }\textbf {\bibinfo {volume} {100}},\ \bibinfo
  {pages} {117201} (\bibinfo {year} {2008})}\BibitemShut {NoStop}%
\bibitem [{\citenamefont {Platt}\ \emph {et~al.}(2001)\citenamefont {Platt},
  \citenamefont {Minor},\ and\ \citenamefont {Klemmer}}]{Platt2001}%
  \BibitemOpen
  \bibfield  {author} {\bibinfo {author} {\bibfnamefont {C.~L.}\ \bibnamefont
  {Platt}}, \bibinfo {author} {\bibfnamefont {M.~K.}\ \bibnamefont {Minor}}, \
  and\ \bibinfo {author} {\bibfnamefont {T.~J.}\ \bibnamefont {Klemmer}},\
  }\href@noop {} {\bibfield  {journal} {\bibinfo  {journal} {IEEE Trans.
  Mag.},\ }\textbf {\bibinfo {volume} {37}},\ \bibinfo {pages} {2302} (\bibinfo
  {year} {2001})}\BibitemShut {NoStop}%
\bibitem [{\citenamefont {Hindmarch}\ \emph {et~al.}(2010)\citenamefont
  {Hindmarch} \emph {et~al.}}]{Hindmarch2010}%
  \BibitemOpen
  \bibfield  {author} {\bibinfo {author} {\bibfnamefont {A.~T.}\ \bibnamefont
  {Hindmarch}} \emph {et~al.},\ }\href@noop {} {\bibfield  {journal} {\bibinfo
  {journal} {Phys. Rev. B},\ }\textbf {\bibinfo {volume} {81}},\ \bibinfo
  {pages} {100407} (\bibinfo {year} {2010})}\BibitemShut {NoStop}%
\bibitem [{\citenamefont {Rushforth}\ \emph {et~al.}(2008)\citenamefont
  {Rushforth} \emph {et~al.}}]{Rushforth2008}%
  \BibitemOpen
  \bibfield  {author} {\bibinfo {author} {\bibfnamefont {A.~W.}\ \bibnamefont
  {Rushforth}} \emph {et~al.},\ }\href@noop {} {\bibfield  {journal} {\bibinfo
  {journal} {Phys. Rev. B},\ }\textbf {\bibinfo {volume} {78}},\ \bibinfo
  {pages} {085314} (\bibinfo {year} {2008})}\BibitemShut {NoStop}%
\bibitem [{\citenamefont {Raanaei}\ \emph {et~al.}(2009)\citenamefont {Raanaei}
  \emph {et~al.}}]{Raanaei2009}%
  \BibitemOpen
  \bibfield  {author} {\bibinfo {author} {\bibfnamefont {H.}~\bibnamefont
  {Raanaei}} \emph {et~al.},\ }\href@noop {} {\bibfield  {journal} {\bibinfo
  {journal} {J. Appl. Phys.},\ }\textbf {\bibinfo {volume} {106}},\ \bibinfo
  {pages} {023918} (\bibinfo {year} {2009})}\BibitemShut {NoStop}%
\bibitem [{\citenamefont {Gonzalez-Guerrero}\ \emph {et~al.}(2007)\citenamefont
  {Gonzalez-Guerrero} \emph {et~al.}}]{Gonzalez-Guerrero2007}%
  \BibitemOpen
  \bibfield  {author} {\bibinfo {author} {\bibfnamefont {M.}~\bibnamefont
  {Gonzalez-Guerrero}} \emph {et~al.},\ }\href@noop {} {\bibfield  {journal}
  {\bibinfo  {journal} {Appl. Phys. Lett.},\ }\textbf {\bibinfo {volume}
  {90}},\ \bibinfo {pages} {162501} (\bibinfo {year} {2007})}\BibitemShut
  {NoStop}%
\bibitem [{\citenamefont {Krebs}\ \emph {et~al.}(1987)\citenamefont {Krebs},
  \citenamefont {Jonker},\ and\ \citenamefont {Prinz}}]{Krebs1987}%
  \BibitemOpen
  \bibfield  {author} {\bibinfo {author} {\bibfnamefont {J.~J.}\ \bibnamefont
  {Krebs}}, \bibinfo {author} {\bibfnamefont {B.~T.}\ \bibnamefont {Jonker}}, \
  and\ \bibinfo {author} {\bibfnamefont {G.~A.}\ \bibnamefont {Prinz}},\
  }\href@noop {} {\bibfield  {journal} {\bibinfo  {journal} {J. Appl. Phys.},\
  }\textbf {\bibinfo {volume} {61}},\ \bibinfo {pages} {2596} (\bibinfo {year}
  {1987})}\BibitemShut {NoStop}%
\bibitem [{\citenamefont {Thomas}\ \emph {et~al.}(2003)\citenamefont {Thomas}
  \emph {et~al.}}]{Thomas2003}%
  \BibitemOpen
  \bibfield  {author} {\bibinfo {author} {\bibfnamefont {O.}~\bibnamefont
  {Thomas}} \emph {et~al.},\ }\href@noop {} {\bibfield  {journal} {\bibinfo
  {journal} {Phys. Rev. Lett},\ }\textbf {\bibinfo {volume} {90}},\ \bibinfo
  {pages} {017205} (\bibinfo {year} {2003})}\BibitemShut {NoStop}%
\bibitem [{\citenamefont {Bianco}\ \emph {et~al.}(2008)\citenamefont {Bianco}
  \emph {et~al.}}]{Bianco2008}%
  \BibitemOpen
  \bibfield  {author} {\bibinfo {author} {\bibfnamefont {F.}~\bibnamefont
  {Bianco}} \emph {et~al.},\ }\href@noop {} {\bibfield  {journal} {\bibinfo
  {journal} {J. Appl. Phys.},\ }\textbf {\bibinfo {volume} {104}},\ \bibinfo
  {pages} {083901} (\bibinfo {year} {2008})}\BibitemShut {NoStop}%
\bibitem [{\citenamefont {Kirk}\ \emph {et~al.}(2009)\citenamefont {Kirk} \emph
  {et~al.}}]{Kirk2009}%
  \BibitemOpen
  \bibfield  {author} {\bibinfo {author} {\bibfnamefont {D.}~\bibnamefont
  {Kirk}} \emph {et~al.},\ }\href@noop {} {\bibfield  {journal} {\bibinfo
  {journal} {Phys. Rev. B},\ }\textbf {\bibinfo {volume} {79}},\ \bibinfo
  {pages} {014203} (\bibinfo {year} {2009})}\BibitemShut {NoStop}%
\bibitem [{\citenamefont {Park}\ \emph {et~al.}(1995)\citenamefont {Park},
  \citenamefont {Fullerton},\ and\ \citenamefont {Bader}}]{Park1995}%
  \BibitemOpen
  \bibfield  {author} {\bibinfo {author} {\bibfnamefont {Y.}~\bibnamefont
  {Park}}, \bibinfo {author} {\bibfnamefont {E.~E.}\ \bibnamefont {Fullerton}},
  \ and\ \bibinfo {author} {\bibfnamefont {S.~D.}\ \bibnamefont {Bader}},\
  }\href@noop {} {\bibfield  {journal} {\bibinfo  {journal} {Appl. Phys.
  Lett.},\ }\textbf {\bibinfo {volume} {66}},\ \bibinfo {pages} {2140}
  (\bibinfo {year} {1995})}\BibitemShut {NoStop}%
\bibitem [{Not()}]{Note1}%
  \BibitemOpen
  \href@noop {} {}\bibinfo {note} {{Methods which assume the reversal is
  single-domain-like~\cite{Bianco2008,Hindmarch2010}\ cannot unambiguously
  determine $K_{\mathrm{U}}^{\mathrm{eff}}$ for these \sevenCoFe\ films: we
  approximate \HK\ as the saturation field along the UHA, resulting in non-zero
  intercept in figure~\ref{figKU}\ as an artefact.}}\BibitemShut {Stop}%
\bibitem [{\citenamefont {Fu}\ and\ \citenamefont {Mansuripur}(1992)}]{Fu1992}%
  \BibitemOpen
  \bibfield  {author} {\bibinfo {author} {\bibfnamefont {H.}~\bibnamefont
  {Fu}}\ and\ \bibinfo {author} {\bibfnamefont {M.}~\bibnamefont
  {Mansuripur}},\ }\href@noop {} {\bibfield  {journal} {\bibinfo  {journal}
  {Phys. Rev. B},\ }\textbf {\bibinfo {volume} {45}},\ \bibinfo {pages} {7188}
  (\bibinfo {year} {1992})}\BibitemShut {NoStop}%
\bibitem [{\citenamefont {Sheng}\ \emph {et~al.}(2006)\citenamefont {Sheng}
  \emph {et~al.}}]{Sheng2006}%
  \BibitemOpen
  \bibfield  {author} {\bibinfo {author} {\bibfnamefont {H.~W.}\ \bibnamefont
  {Sheng}} \emph {et~al.},\ }\href@noop {} {\bibfield  {journal} {\bibinfo
  {journal} {Nature},\ }\textbf {\bibinfo {volume} {439}},\ \bibinfo {pages}
  {419} (\bibinfo {year} {2006})}\BibitemShut {NoStop}%
\bibitem [{\citenamefont {Hirata}\ \emph {et~al.}(2011)\citenamefont {Hirata}
  \emph {et~al.}}]{Hirata2011}%
  \BibitemOpen
  \bibfield  {author} {\bibinfo {author} {\bibfnamefont {A.}~\bibnamefont
  {Hirata}} \emph {et~al.},\ }\href@noop {} {\bibfield  {journal} {\bibinfo
  {journal} {Nat. Mater.},\ }\textbf {\bibinfo {volume} {10}},\ \bibinfo
  {pages} {28} (\bibinfo {year} {2011})}\BibitemShut {NoStop}%
\bibitem [{\citenamefont {Srivastava}(1977)}]{Srivastava1977}%
  \BibitemOpen
  \bibfield  {author} {\bibinfo {author} {\bibfnamefont {R.~S.}\ \bibnamefont
  {Srivastava}},\ }\href@noop {} {\bibfield  {journal} {\bibinfo  {journal} {J.
  Appl. Phys.},\ }\textbf {\bibinfo {volume} {48}},\ \bibinfo {pages} {1355}
  (\bibinfo {year} {1977})}\BibitemShut {NoStop}%
\bibitem [{\citenamefont {Robinson}(1962)}]{Robinson1962}%
  \BibitemOpen
  \bibfield  {author} {\bibinfo {author} {\bibfnamefont {G.}~\bibnamefont
  {Robinson}},\ }\href@noop {} {\bibfield  {journal} {\bibinfo  {journal} {J.
  Phys. Soc. Jpn. Suppl.},\ }\textbf {\bibinfo {volume} {17}},\ \bibinfo
  {pages} {558} (\bibinfo {year} {1962})}\BibitemShut {NoStop}%
\bibitem [{\citenamefont {West}(1964)}]{West1964}%
  \BibitemOpen
  \bibfield  {author} {\bibinfo {author} {\bibfnamefont {F.~G.}\ \bibnamefont
  {West}},\ }\href@noop {} {\bibfield  {journal} {\bibinfo  {journal} {J. Appl.
  Phys.},\ }\textbf {\bibinfo {volume} {35}},\ \bibinfo {pages} {1827}
  (\bibinfo {year} {1964})}\BibitemShut {NoStop}%
\bibitem [{\citenamefont {Barandiaran}\ \emph {et~al.}(2003)\citenamefont
  {Barandiaran} \emph {et~al.}}]{Barandiaran2003}%
  \BibitemOpen
  \bibfield  {author} {\bibinfo {author} {\bibfnamefont {M.}~\bibnamefont
  {Barandiaran}} \emph {et~al.},\ }\href@noop {} {\bibfield  {journal}
  {\bibinfo  {journal} {J. Non-Cryst. Solids},\ }\textbf {\bibinfo {volume}
  {329}},\ \bibinfo {pages} {194} (\bibinfo {year} {2003})}\BibitemShut
  {NoStop}%
\bibitem [{\citenamefont {Hashimoto}\ \emph {et~al.}(2008)\citenamefont
  {Hashimoto} \emph {et~al.}}]{Hashimoto2008}%
  \BibitemOpen
  \bibfield  {author} {\bibinfo {author} {\bibfnamefont {A.}~\bibnamefont
  {Hashimoto}} \emph {et~al.},\ }\href@noop {} {\bibfield  {journal} {\bibinfo
  {journal} {IEEE Trans. Mag.},\ }\textbf {\bibinfo {volume} {44}},\ \bibinfo
  {pages} {3899} (\bibinfo {year} {2008})}\BibitemShut {NoStop}%
\end{thebibliography}


%

\end{document}